# HUMAN-CENTERED HUMAN-AI COLLABORATION (HCHAC)

A. Author's name and details

Qi Gao (Zhejiang University, Hangzhou, China, qi.gao@zju.edu.cn, https://orcid.org/0000-0002-4984-877X),

Wei Xu (Zhejiang University, Hangzhou, China, weixu6@yahoo.com, https://orcid.org/0000-0001-8913-2672),

Hanxi Pan (Zhejiang University, Hangzhou, China, hanxipan@zju.edu.cn, https://orcid.org/0000-0003-0415-0471),

Mowei Shen (Zhejiang University, Hangzhou, China, mwshen@zju.edu.cn, https://orcid.org/0000-0001-7661-2968)

Zaifeng Gao (Zhejiang University, Hangzhou, China, zaifengg@zju.edu.cn, https://orcid.org/0000-0001-9727-8524)

Corresponding Author: Zaifeng Gao

B. Abstract

In the intelligent era, the interaction between humans and intelligent systems fundamentally involves collaboration with autonomous intelligent agents. Human-AI Collaboration (HAC) represents a novel type of human-machine relationship facilitated by autonomous intelligent machines equipped with AI technologies. In this paradigm, AI agents serve not only as auxiliary tools but also as active teammates, partnering with humans to accomplish tasks collaboratively. Human-centered AI (HCAI) emphasizes that humans play critical leadership roles in the collaboration. This human-led collaboration imparts new dimensions to the human-machine relationship, necessitating innovative research perspectives, paradigms, and agenda to address the unique challenges posed by HAC.

This chapter delves into the essence of HAC from the human-centered perspective, outlining its core concepts and distinguishing features. It reviews the current research methodologies and research agenda within the HAC field from the HCAI perspective, highlighting advancements and ongoing studies. Furthermore, a framework for human-centered HAC (HCHAC) is proposed by integrating these reviews and analyses. A case study of HAC in the context of autonomous vehicles is provided, illustrating practical applications and the synergistic interactions between humans and AI agents. Finally, it identifies potential future research directions aimed at enhancing the effectiveness, reliability, and ethical integration of human-centered HAC systems in diverse domains.

# 1. Introduction

In 2015, Google's AI system AlphaGo marked the beginning of an intelligent era built on AI, machine learning, big data, and cloud computing technologies. The emergence of GPT-4 in 2023 further accelerated this transformation. While earlier intelligent technologies primarily operated in specialized domains like industrial robotics and manufacturing—areas relatively removed from everyday life—today's intelligent systems have seamlessly integrated into daily routines. Modern intelligent agents (robots, virtual assistants, and autonomous vehicles) demonstrate not only human-like cognitive abilities such as learning and reasoning but also emotional and social capabilities including empathy and social etiquette, establishing a new level of autonomy. This autonomy enables them to independently analyze unexpected situations, develop novel solutions beyond predefined parameters, and accomplish tasks that previous automation technologies could not address (Kaber, 2018; Huang, 2024).

The relationship between humans and intelligent agents has consequently evolved from a unidirectional interaction where humans simply command machines to a bidirectional partnership characterized by human-AI collaboration (HAC). This transition has heightened awareness about prioritizing human needs and values in AI development. Building on this new collaborative relationship, human-centered and human-led AI approaches aim to ensure that AI systems enhance human capabilities rather than replace them, promoting utility and human values throughout the AI lifecycle.

This chapter provides a comprehensive examination of HAC from a human-centered perspective, addressing both theoretical foundations and empirical evidence in this emerging field. The levels and forms of human-AI interaction are first defined, with a distinction drawn between HAC and automation or related paradigms. Subsequently, research questions that conceptualize AI as teammates within collaborative cognitive systems under human leadership are examined, with attention to human factors across four key team dimensions: cognition, control, transaction, and relationship. Based on these frameworks, a model of human-led AI collaboration is proposed, underscoring the critical role of human leadership in future HAC research. A case study in autonomous vehicles is used to illustrate the complexities and challenges inherent in human-led HAC systems. The chapter concludes by outlining future research directions, emphasizing the necessity of continued development of frameworks to enhance the effectiveness, reliability, and ethical integration of HAC.

## 2. An Overview of Human-AI Collaboration

### 2.1. Defining Different Levels of Human-AI Collaboration

In HAC, machines have moved beyond serving solely as auxiliary tools in the mechanical and information eras to become true collaborators and teammates. This dual role of intelligent agents as "auxiliary tools + collaborative teammates" has become a defining characteristic of autonomous systems in the intelligent era (Committee et al., 2022). The 'teammate' design metaphor for AI poses potential risks, including the possibility of humans losing control over AI systems (Shneiderman, 2022). To address the challenges, Xu and Gao (2024) advocate that humans must maintain a leadership role in HAC to ensure effective control. Such "human-AI teaming" endows the human-machine relationship with new meaning, ushering in an evolutionary leap in how humans and machines interact. Consequently, researchers are now called upon to explore novel theories and methods to study human intelligence collaboration within new research paradigms.

Different levels of HAC can be understood through a progressive framework of AI capabilities (see Table 1). While traditional automation systems (L0-L1) can only perform predetermined or rule-based actions without true decision-making abilities, modern AI agents range from machine learning systems (L2) to sophisticated collaborative partners (L5), with increased human-like capabilities. As soon as AI systems acquire reasoning and decision-making abilities (L2), they can operate autonomously. When memory, reflection, learning, and generalization capacities (L3–L4) are incorporated, they learn from experience, predict consequences, adapt to new circumstances, and refine their behavior based on accumulated knowledge. For instance, advanced autonomous driving systems can learn driver preferences, continuously update safety protocols, and tailor driving behaviors accordingly. Indeed, current autonomous systems demonstrate far greater independence, initiative, and autonomy than traditional automation (O'Neill et al., 2022). However, as AI becomes increasingly human-like, there is a need to cultivate social and collaborative characteristics (L5) to match or even surpass the performance of all-human teams.

Table 1 Levels of AI (Huang, 2024)

| AI Agent Levels | Techniques & Capabilities |
|---|---|
| L0: | No AI + Tools (Perception + Actions) |
| L1: | Rule-based AI + Tools (Perception + Actions) |
| L2: | Imitation Learning/Reinforcement Learning-based AI + Tools (Perception + Actions) + **Reasoning & Decision Making** |
| L3: | Large-Language-Model (LLM) -based AI + Tools (Perception + Actions) + Reasoning & Decision Making + **Memory & Reflection** |
| L4: | LLM-based AI + Tools (Perception) + Actions + Reasoning & Decision Making + Memory & Reflection + **Autonomous Learning + Generalization** |
| L5: | LLM-based AI + Tools (Perception) + Actions + Reasoning & Decision Making + Memory + Reflection + Autonomous Learning + Generalization + **Personality (Emotion + Character) + Collaborative behavior (Multi-Agents)** |

*Note.* **Bold** font indicates the human-like abilities that AI gains at this level compared to the previous one.

This AI-capability-driven approach offers a concise perspective for classifying levels of HAC. However, it reflects a technology-centered viewpoint, the undergoing HAC research needs a shift toward a human-centered perspective (Xu & Gao, 2024). This shift calls for diving into how humans and AI collaborate at each level. While higher-level AI agents possess significant autonomy and can modify operational goals as needed, they should not have final decision-making authority. The power to define fundamental work goals and make critical decisions remains with human operators. In HAC, even the most advanced AI agents (L5) are designed to augment and empower human capabilities rather than replace human agency, remaining aligned with human objectives and under human oversight.

2.2. Human-AI Collaboration, Human-AI Teaming, and Human-AI Hybrid Intelligence and Behavior

As a new paradigm of human-machine interaction that emerged in the era of autonomous intelligence, HAC refers to the continuous joint work between humans and AI (or autonomous intelligent agents) toward common goals. According to Lee et al. (2023), the highest level of human-AI teaming—collaboration—builds on two foundational levels: (1) Coordination, where humans and AI agents synchronize the timing of tasks and resources to optimize performance, and (2) Cooperation, where humans and AI agents use negotiation, alongside coordination, to resolve conflicts between individual and collective goals. Through these lower levels, collaboration evolves, where humans and AI agents share authority, make joint decisions over extended periods, and develop shared rules, norms, and values to address dynamic and often unpredictable challenges. The hierarchy of coordination, cooperation, and collaboration, spans a spectrum from low to high resilience. This resilience refers to a system's capacity to not only withstand unexpected disruptions but also to thrive in the face of dynamically shifting environments—all while ensuring continued effective performance. Furthermore, as systems progress from mere coordination to cooperation and ultimately to collaboration, the time constant property—defined as the duration over which interactions and relationships develop and mature—lengthens. This elongation of the time scale is indicative of the increasing complexity and depth of the relationships within the system, which are necessary to support more sophisticated forms of collaboration.

The collaborative level of human-AI teaming is characterized by two extra features, dynamic and bidirectional, compared with broad human-computer interaction (HCI). O'Neill et al. (2022) proposed that possessing the following characteristics is enough for constituting human-AI teaming: (1) The intelligent agent is regarded as an "independent entity" by human teammates in the human-machine team, and the intelligent agent has a considerable degree of independent decision-making ability; (2) The role of the intelligent agent must be interdependent with the role of the human teammates; (3) There must be one or more humans and one or more intelligent agents forming a human-machine team. However, for HAC, the focus of human-AI teaming shifts from predefined roles to a horizontal partnership, where both agents adapt dynamically to evolving situations, characterizing the dynamic feature. For the bidirectional feature, both human and AI agents actively communicate, exchange ideas, and jointly make decisions. This high-level integration ensures both human and AI agents develop long-term trust, shared understanding, and mutual adaptability, critical for handling complex, real-world task scenarios.

While traditional human-machine interaction might simply combine human and machine capabilities additively, HAC aims to achieve hybrid (fusion) intelligence and behavior, forming complementary advantages between humans and machines to develop more powerful and sustainable enhanced teams. Hybrid-enhanced intelligence enables collaborative decision-making in complex problems, achieving capabilities beyond what either humans or AI agents could accomplish independently (Crandall et al., 2018). This can be implemented through "human in the loop" or "brain in the loop" collaborative systems (Xu et al., 2021). In human-in-the-loop collaboration, when the AI agent has low confidence in its output, humans proactively intervene to guide the solution, creating a feedback loop that enhances overall system intelligence. Brain-in-loop collaboration aims for deeper integration of biological and machine intelligence through neural connection channels, enabling enhancement, replacement, or compensation of certain AI agent functions while maintaining meaningful human-AI collaboration. Furthermore, these hybrid human-machine systems form complex socio-technical interactions where machines can shape human behaviors (e.g., through news filtering, social media interactions, or learning outcomes) while humans shape machine behaviors through engineering choices, training data, and feedback processes (Rahwan et al., 2019).

2.3. Comparing Human-Automation, Human-Autonomy, and Human-Human Interaction

Earlier in this chapter, the HAC levels from the perspective of AI capabilities were discussed. Another useful categorization approach is based on the collaboration extents. HAC research always falls within the broader domain of human-autonomy interaction. Beyond this domain, as the collaboration weakens, HAC shifts toward human-automation interaction, where the role of AI is more passive and task-oriented. Conversely, as the collaboration strengthens, HAC moves toward human-human interaction, where AI plays a more active role.

Comparison between autonomous and automated systems has long been recognized with the 10 levels of automation (LOA) continuum proposed by Sheridan and Verplank (1978). O'Neill et al. (2022) extended this framework by dividing the levels of autonomy based on the LOA, highlighting how the role of the intelligent agent in HAC changes as the LOA increases (Table 2). At lower levels (LOA 1-4), the machine only provides information about the task, lacking the freedom to make decisions and the ability to independently engage in activities without pre-programmed instructions. As it moves to higher levels (LOA 5-6), the machine begins to recommend and execute actions, albeit with human oversight, such as in collision warning systems. At the highest levels (LOA 7-10), the machine operates autonomously, executing actions without human intervention, achieving the highest level of autonomy. The transition from automated to autonomous systems can be characterized by the degree of control or agency the AI agent possesses. At the highest levels, the agent makes decisions and controls independently, while at lower levels, it may only provide suggestions or require human approval. Shneiderman (2022) criticizes this traditional one-dimensional automation/autonomy framework—which assumes that increased automation necessarily leads to reduced human involvement—as being oversimplified, outdated, and potentially dangerous for designing trustworthy AI systems. Instead, Shneiderman (2022) offers a two-dimensional framework of automation and human control, a structured way to evaluate and design human-centered AI systems by balancing *AI autonomy* and *human control*. It addresses the concern that increasing autonomy may diminish meaningful human oversight in order to achieve reliable, safe, and trustworthy AI.

Table 2 The 10 levels of automation and the corresponding levels of autonomy (LOA, O'Neil et al., 2022)

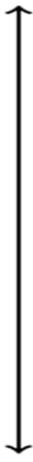

| LOA | Automation Level | Agent Autonomy | Automation or Autonomous Agent Role and Capability |
|---|---|---|---|
| 10 | High | High agent autonomy | The computer decides everything and acts autonomously, ignoring the human |
| 9 | | | The computer informs the human only if it, the computer, decides to |
| 8 | | | The computer informs the human only if asked, or |
| 7 | | | The computer executes automatically, then necessarily informs the human, and |
| 6 | | Partial agent autonomy | The computer allows the human a restricted time to veto before automatic execution, or |
| 5 | | | The computer executes that suggestion if the human approves, or |
| 4 | | No autonomy / Manual control | The computer suggests one alternative, or |
| 3 | | | The computer narrows the selection down to a few, or |
| 2 | | | The computer offers a complete set of decision/action alternatives, or |
| 1 | | | The computer offers no assistance; the human must take all decisions and actions. |
| | Low | | |

As levels of AI improved, research in HAC has drawn parallels with human-human interaction, where a broader viewpoint based on human-team-related concepts was adopted. For example, Iftikhar et al. (2024) directly extended a human-team research framework into HAC to cater to the dynamic, multilevel, and complex view of the teams in HAC. This framework conceives team structural factors, compositional factors, and mediating mechanisms as overlapping and coevolving aspects in human-AI teams. For human teams, verbal or non-verbal communication plays a crucial role in mediating information sharing depending on each other's joint attention, perspective taking, consensus building, and mental state and behavior inferencing (Krämer et al., 2012). These key abilities have been already gaining attention from current AI agents (e.g., Belkaid et al., 2021; Yuan et al., 2022). At the same time, human teams gradually form a sense of belonging, social reciprocity, trust, and equality, which also appear in the interaction between humans and intelligent agents (Lyons et al., 2021).

However, HAC has distinct advantages and disadvantages, which determine the need for new research paradigms stepping out of the shadow of human teams. First, HAC realizes hybrid augmentation, where humans and AI can complement each other in cognition and behavior, breaking through task boundaries that traditional human teams cannot handle, e.g., surgical robots diagnosing in-vivo status. Second, HAC can cooperate in a more coordinated and harmonious way, with the adaptation of AI agents to humans being determined by humans, offering personalized and flexible advantages, leading to high-quality team cooperation (Salas & Fiore, 2004). Furthermore, HAC may reduce the ambiguity and vagueness brought about by social interaction in human teams, as well as potential task obstacles due to social relationships and hierarchies. In this way, research should highlight both the uniqueness and commonalities of HAC and human-human teams. However, HAC should not be viewed simply as an imitation of human teams, nor should mimicking human-human teams be seen as the end goal (McNeese et al., 2023).

Currently, HAC remains constrained by the advancement of AI-related technologies, and also insufficient exploration of team mechanisms including information processing and response methods of collaborating with intelligent agents compared to humans. A meta-analysis by Vaccaro et al. (2024) revealed that, on average, HAC did not outperform the best of humans or AI working alone, suggesting that while HAC holds potential, it is not universally superior. Demir et al. (2017) found that human-AI team performance can reach a level comparable to that of non-expert human teams but cannot match expert human teams. The differences in knowledge, experience, and problem-solving strategies between humans and intelligent agents may lead to cooperation bottlenecks in HAC, especially when unforeseen events occur, as the response or processing speed of HAC to changes is slower than that of human teams (Lyons et al., 2021). Notably, performance gains in HAC were observed in content creation tasks, outperforming the best individual component, whereas similar improvements were not as pronounced in decision-making tasks. This suggests that HAC has significant potential for fostering human-AI synergy, particularly in areas like content creation (Vaccaro et al., 2024).

2.4. HAC Research Orientation

Through the systematic literature search and content analysis, Berretta et al. (2023) identified five clusters describing the research orientation in the HAC field (Table 3): 1) A human-oriented cluster, rooted in psychology, that examines human factors such as preferences, trust, and situation awareness; 2) A task-oriented cluster, rooted in information science, that discusses different AI roles and collaboration strategies based on task context; 3) An explainability cluster that focuses on how AI explainability can enhance decision-making process, calibrated trust and HAC; 4) A technology-oriented cluster that centers on human-robot collaboration, emphasizing security aspects through communication; and 5) An agent-oriented cluster that explores the development of human trust in AI based on performance and the shift towards AI as a collaborative partner. These clusters highlight the diverse aspects of HAC, from human factors, and design science, to computer science and engineering, which requires multi-discipline collaborations.

Table 3 Five clusters identified in Human-AI teaming-related research (adapted from Berretta et al., 2023)

| Perspective | Human-oriented | Task-oriented | Explainability-oriented | Robot-oriented | Agent-oriented |
|---|---|---|---|---|---|
| Methods | Mainly mixed methods, qualitative interviews, field and online experiments, literature review | Mixed methods, vignette study, theory and framework development, literature synthesis, commentary, experiment, experience sampling | Mixed-methods, laboratory experiments with Wizard of OZ or real AI | Mainly theory and framework development, literature review, partly enriched with exemplary studies | Laboratory and online experiments, panel invitation |
| Forms of AI | Decision(support) system, variety of software or embodied agents | Decision(support) system, robot | Decision(support) system, virtual drone | Robots with machine or reinforcement learning techniques | Embodied agents, software |
| Role and understanding of AI | Different roles from decision support to mutually supporting team member, augmentor of intelligence, support in decision-making, a full, active member with an own role, equal partner, social counterpart | Different roles, augmentor of intelligence, decision agent, independent, active agent, partner & teammate | Assistant & helper, advisor | Autonomous agent, (physical) interaction partner | Autonomous agent, conversational agent, partner, teammate rather than tool |
| Terms for human-AI teaming | Cooperation, human-AI collaboration, human-autonomy-teaming, human-machine team, human-AI teaming | Human-AI cooperation, collaboration, augmented intelligence, human-computer symbiosis | Human-AI collaboration, collaborative partnership, algorithm-in-the-loop, AI-assisted decision-making, human-AI partnership, human-agent/robot/drone teaming | Human-robot interaction, human-robot collaboration, duality, human-robot team | Mixed-initiative interactions, human-AI collaboration, human-AI teaming, autonomy as teammate |
| Understanding of human-AI teaming | Independent agents working toward a common goal, adaptive roles within the team | Differentiated understanding from independent to interdependent, integrated architecture | Spectrum from full automation to full human agency, AI assistance to support humans | Supportive relationship, working together for task accomplishment, co-working with the influence of each's values, and broadening individual capabilities | Complementary strengths, prompt interaction in response to communication flow, adaptation toward dynamic tasks to achieve the common goal |
| Contexts under examination | Hospitality, production management, cyber incident response, sequential risky, decision-making, context-free | Context-aware services, managerial decision-making, financial markets, gig economy platforms, autonomous driving | Clinical decision-making, user experience, content moderation, performance prediction, gaming | Manufacturing, production, industry, safety, context-free | Design, military, autonomous driving, context-free |

For the human-AI interaction field, the research focus shifts from traditional HCI paradigms significantly as it evolves to accommodate the complexities of HAC. Table 4 outlines eight critical transitions that characterize this evolution, ranging from changes in machine behavior predictability to the emergence of metaverse interactions (Xu et al., 2024). These transitions reflect how the field has moved beyond simple user interface design and usability concerns to encompass broader challenges such as ethical AI development, human-controllable autonomy, and hybrid intelligence systems. The framework presents a comprehensive overview of how research priorities have shifted to address new complexities in human-AI interaction while highlighting the emerging focus areas in human factors research that are essential for developing effective, ethical, and human-centered AI systems.

Table 4 Research Focus Transition in Human-AI Interaction (Xu et al., 2024)

| New Features | New Issues | Research Focus |
| --- | --- | --- |
| From "expected" to "potentially unexpected" machine behavior | Intelligent systems may exhibit uncertain behaviors and unique evolutions, leading to system output biases; Existing software testing methods lack consideration for intelligent machine behavior; Intelligent systems possess characteristics such as evolution and social interaction | Behavioral science methods for studying machine behavior; Iterative design and user testing methods to reduce system output biases during data collection, training, and algorithm testing; User-participatory design, "human-centered" machine learning |
| From "human-computer interaction" to "human-intelligence team collaboration" | Machines (intelligent agents) transition from tools to teammates collaborating with humans; How to model human-machine collaboration (shared situational awareness, psychological models, decision-making, and control) | Theories and methods based on human-intelligence team collaboration paradigms; Hac theories, models, and team performance evaluation systems |
| From "human intelligence" to "human-machine intelligence complementary" hybrid augmented intelligence | Machines cannot replicate advanced human cognitive abilities, and the isolated development of machine intelligence encounters bottlenecks; It is uncertain how to integrate human roles into intelligent systems for collaborative synergy | Cognitive architecture for human-machine hybrid augmented intelligence; Collaborative human-multiple intelligent agent systems based on the paradigm of collaborative cognitive ecosystems |
| From "human-centered automation" to "human-controllable autonomy" | Potential loss of human control over autonomous intelligent systems; Potential negative impacts of automation technology (uncertainty, etc.) | Paradigms for HCI design for autonomous technology; Human factors methods for human-controlled autonomy; shared autonomy control design between humans and machines |
| From "non-intelligent" to "intelligent" human-machine interaction | Enhancing the naturalness of intelligent user interfaces; Effective design of intelligent human-machine interaction; Bottleneck effects of human perceptual and cognitive resources in ubiquitous computing environments | New paradigms for intelligent HCI and interface design; Multi-channel natural user interface design; New types of HCI technologies and designs (affective interaction, intent recognition, brain-computer interfaces, etc.); Human factors design standards for intelligent technology research and development |
| From "experience demand" to "ethical AI" | New user demands (privacy, ethics, fairness, skill development, decision-making rights, etc.); Potential output biases and unexpected results of intelligent systems; Misuse of intelligent systems (discrimination, privacy issues, etc.); Lack of traceability and accountability mechanisms for intelligent system failures | Interdisciplinary design of ethical AI based on human factors methods; methods based on the paradigm of collaborative cognitive ecosystems; methods based on the paradigm of intelligent socio-technical systems; Meaningful human control; transparent design |
| From "experiential" to "systematic" interaction design | Limitations of user experience and usability design methods; How to effectively carry out prototyping and usability testing of intelligent systems; Human factors personnel have not been involved early in the development of intelligent systems | The development process of intelligent systems based on human factors concepts; User experience-driven intelligent innovation design; effective intelligent interaction design methods; Systematic human factors methods |
| From "physical entity" to "virtual-real integrated metaverse" space interaction | New demands for HCI in virtual-real integrated spaces; New experiences of immersion, interactivity, and parallel seasonal environments in the metaverse; Ethics, information presentation, brain-computer integration, etc., in the metaverse interaction space; New issues brought by multimodal continuity and ambiguity of interaction data for interaction intent inference in the metaverse space | Natural HCI patterns and technologies in the metaverse space; Virtualization, remoteness, and multi-mapping relationships in HCI; Social relationships between people and between people and intelligence in the metaverse interaction space |

To reflect the focus transition in HAC, Xu and Gao (2024) introduced the conceptual model for HAC. This model encapsulates the human-AI interaction system by integrating the joint cognitive systems theory (Hollnagel & Woods, 2005), situation awareness (SA) theory (Endsley, 1988), and intelligent agent theory (Wooldridge & Jennings, 1995; see Figure 1). This model conceptualizes the HAC as a joint cognitive system, wherein AI agents possess cognitive processing capabilities analogous to humans. In this framework, the relationship between humans and AI agents is interdependent. Through the human-AI cognitive interface, both parties are capable of establishing shared SA and aligned goals, fostering mutual trust, and comprehensively understanding each other's states and contextual information, while maintaining the ultimate authority of humans. The model employs a heterogeneous homomorphic approach to depict the information processing within the system. Drawing from SA theory, human cognition of external environmental information involves perception, comprehension, and prediction processes (Figure 1, left). The intelligent system is similarly structured into three layers of sensing, reasoning, and forecasting (Figure 1, right). This model introduces seven novel insights, offering a comprehensive framework and innovative perspectives for advancing HAC research.

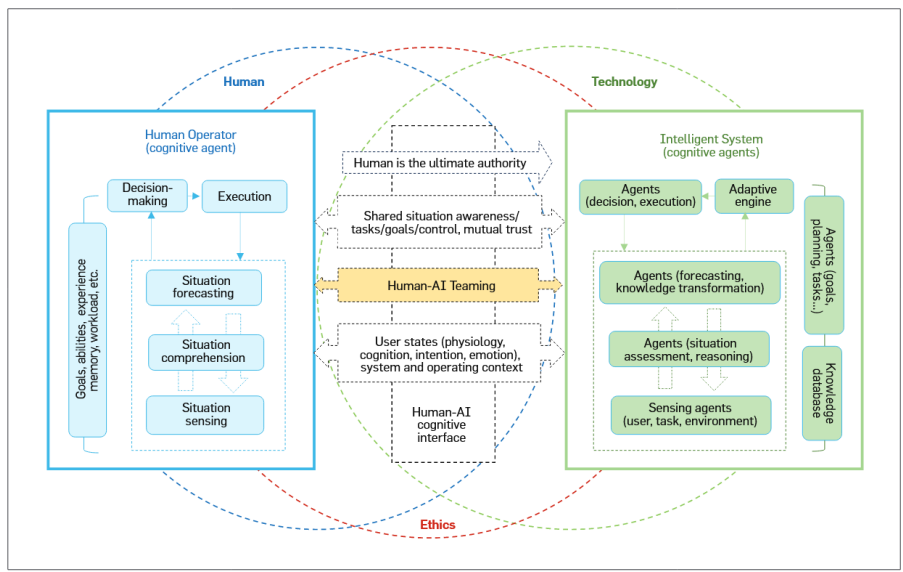

Figure 1 The Conceptual Model for HAC

(1) New research paradigm based on HAC. Unlike the unidirectional human-machine interactions characteristic of traditional ergonomics, the HAC model redefines the human-intelligence relationship as a collaborative partnership within a joint cognitive system, emphasizing the enhancement of the overall performance of the team by optimizing collaborative interactions.

(2) New research approach based on machine cognitive entities. Departing from conventional ergonomics research, which typically regards machines as mere tools to support human operations, this approach conceptualizes machine intelligence entities as cognitive partners that actively team up with humans. This perspective allows researchers to investigate ways to improve the performance of HAC by examining and optimizing the cognitive behaviors of intelligent agents and their interactions with humans.

(3) Persist and expand the "human-centered AI" approach. Intelligent systems must identify optimal control switch points by accurately perceiving and recognizing user states and situational contexts, thereby improving the system's risk response capability and team performance. In emergencies, human operators retain ultimate decision-making authority, ensuring that human users maintain a dominant role and preventing potential conflicts or losses arising from operational disagreements between users and intelligent systems.

(4) Human-AI bidirectional proactive recognition. Unlike the traditional "stimulus-response" one-way human-machine interaction, this model emphasizes human-intelligence bidirectional proactive state recognition. Intelligent entities actively monitor and recognize user physiological, cognitive, behavioral, intentional, emotional, and other states through sensing systems, while humans obtain the best SA through multimodal interfaces.

(5) Emphasize the complementarity of humans and machines. The bidirectional interaction framework enables humans and intelligent agents to complement each other's strengths. As components of a collaborative cognitive system, system performance relies not only on individual capabilities but also on the complementary and cooperative integration of human and machine intelligence and behavior. This synergy maximizes collaborative efficiency and overall team performance.

(6) Adaptability and co-evolution as an inherent requirement of HAC. Adaptability refers to the ability of intelligent agents to adjust their behaviors and strategies in real time based on changing user needs, contexts, and emerging challenges. Co-evolution emphasizes the mutual development and refinement of both human and AI capabilities, fostering a synergistic relationship where each party learns from and enhances the other. This dynamic interplay enables HAC to maintain resilience and efficiency, even as tasks and environments become increasingly complex.

(7) Human-machine cooperative cognitive interface. To achieve effective human-intelligence collaboration, the model emphasizes the development of a cooperative cognitive interface grounded in multimodal interaction technologies. Such interfaces support various aspects, including bidirectional SA, mutual trust, shared decision-making, shared control, social interaction, and emotional interaction between humans and machines.

### 3. Human Factor Research Methods of HAC

3.1. Research Paradigm

The human factors research domain has undergone several research paradigm transitions from the mechanical era to the computer era and intelligent era. Xu et al. (2024) outlined these transitions from the most micro perspective based on individual cognitive neurosciences to the most macro perspective based on the collaboration within the socio-technical system (Table 5). The early paradigm in the computer era continues in parts of modern HAC, where machines were conceptualized as tools assisting humans, optimizing interaction and usability to adapt machines to human needs. In the intelligent era, the dominant paradigm views humans and AI as collaborative cognitive agents within unified systems, emphasizing team-based cognitive collaboration to optimize overall performance. Additionally, there are emerging paradigms just beginning to take shape. One of these envisions collaboration at a broader scale, involving ecosystems of intelligent agents, and the integration of social, organizational, and intelligent subsystems, highlighting interdisciplinary approaches. Another paradigm grounded in human cognitive neuroscience, focusing on the neural mechanisms of cognition in human-computer environments, is also gaining traction, laying the foundation for neural interface technologies and advanced cognitive modeling.

Table 5 Research Paradigm Transition in Human Factors (Xu et al., 2024)

| Era of Emergence | Research Paradigm Orientation | Description of Paradigm Orientation | Representative Domains or Frameworks | Representative Methods |
|---|---|---|---|---|
| Computer and intelligent era | Based on human cognitive neuroscience | Understand the relationship between cognitive neural mechanisms and work performance in human-computer environments | Cognitive neuroscience | Brain-computer interface technology and design, EEG measurement, feature analysis and modeling |
| Computer era | Based on human cognitive information processing | Understand the relationship between cognition and work performance in human-machine interaction at the level of human mental activities, such as perception, memory, and cognition | Engineering psychology | In human-computer operating environments, using work performance measurement (reaction time, error rate, etc.) and subjective evaluation methods to evaluate the relationship between human psychological activities and performance, and to optimize human-computer system design |
| Mechanical era | Based on the differences and complementarity of human and machine capabilities | Optimize the allocation of human and machine functions and tasks, and humans adapt to machines | Early ergonomics, human factors | Human physical task analysis, time-motion analysis, human-machine function and task analysis and allocation, etc. |
| Computer era | Human-computer interaction using machines as assistants (tools) for human tasks | HCI technology, design, testing, and HCI to achieve machine adaptation to humans, HCI optimization, and best user experience | Human-computer interaction | User psychological modeling needs research and analysis, HCI cognitive modeling, and interface conceptualization, based on psychological methods for usability testing |
| Intelligent era | Based on the collaboration between humans and AI both as cognitive agents | The AI agent becomes a team member collaborating with humans. Humans and AI agents are two cognitive entities within a collaborative cognitive system, achieving optimal overall system performance through teamwork and cooperation | Collaborative cognition system | Based on theories such as human-human team collaboration, collaborative cognition system, etc., modeling bidirectional SA, mental models, trust, and decision-making, to achieve human-computer collaboration |
| Intelligent era | Based on human-intelligent networks (multiple collaborative cognitive systems) | Collaboration between multiple AI agents (collaborative cognitive systems) forms an intelligent collaborative cognitive ecosystem, and the overall system performance depends on the collaboration between collaborative cognitive systems | Intelligent collaborative ecosystem | Based on the modeling, design, and technology of the ecosystem, including collaboration between multiple intelligent systems, group knowledge transfer, self-organization and adaptation, distributed situational awareness, HCI, and collaborative decision-making |
| Intelligent era | Based on the collaboration between social and intelligent subsystems | Through the realization of interaction and collaboration between human, organization, society, and technology subsystems, to achieve optimal overall system performance | Intelligent socio-technical system | Systematization methods, social-technical system methods, work system redesign, organizational optimization design, engineering, social behavioral science, and other interdisciplinary methods |

Building upon the research paradigm from HCI, research in HAC has evolved beyond the traditional development framework. This evolution encompasses three primary stages—requirement analysis, design modeling, and testing evaluation—each specifically tailored to address the novel features inherent in HAC systems. The initial phase of requirement analysis necessitates a thorough analysis of tasks and human needs. In traditional HCI research, machine behaviors are typically predefined based on fixed user research results. However, HAC application scenarios are often of high uncertainty, with user needs and states frequently changing. Consequently, the new research paradigm must leverage online real-time data to develop more accurate application profiles including the modeling of user states and task contexts. This approach enables the dynamic alignment of personalized needs for human operators (Berndt et al., 2017). Additionally, HAC researchers must account for unforeseen scenarios by analyzing factors that may constrain human decision-making within tasks, thereby establishing models that facilitate system adaptability (Vicente, 1999). During the promotion and implementation phases, AI technologies may be influenced by various social factors. It is essential to incorporate considerations of privacy, ethics, and other socio-moral elements to develop intelligent systems that are harmonious with the social environment (Xu et al, 2024).

In the design phase of intelligent systems, three interaction orientations are considered: human-to-AI interaction, AI-to-human interaction, and bidirectional interaction. Human-to-AI interaction relies on various input devices. Beyond traditional key inputs, natural interaction methods such as voice, gesture, and brain-computer interfaces have emerged in HAC. Related research often involves collecting data through laboratory experiments to adjust algorithms or interaction parameters. AI-to-human interaction is grounded in system behavior and user interface design. Traditional user interface research typically emphasizes visual elements. However, the increasing complexity and the essence of the cognitive entity of AI agents necessitates that their behavior and multimodal communication serve as means for human teammates to understand them. Consequently, providing more transparent and explainable AI has become a crucial research focus (Gunning et al., 2019). Efficient bidirectional interaction depends on function allocation, workflow, and workspace design. Unlike traditional fixed task allocation, HAC emphasizes the "AI first" concept, prioritizing the use of the machine's intelligence within the work process to reduce repetitive human activities and only assigning tasks to humans when intelligent agents are unable to complete them (Xu, 2022). Currently, most related research is conducted in laboratory settings, utilizing abstract tasks or simulation technologies—such as autonomous driving simulators—to transform design challenges into manipulable independent and dependent variables for experimental studies. Beyond inferential statistics, advanced quantifying methods are adopted to depict the team processes, such as recurrence quantification analysis, pairwise cross-correlations, cluster phase synchrony, pattern entropy, and multivariate omega complexity (Wiltshire et al., 2024).

After designing new interfaces and establishing new models, researchers will test the feasibility of HAC. In the stage where there is only a design prototype and no functional implementation, the Wizard of Oz (WOZ) is one of the typical methods used in HAC research, that is, letting humans play the role of "intelligent machines" to interact with users, to simulate and verify the entire HAC design idea (Martelaro & Ju, 2017). When both the algorithm and interface functions are perfect, it is possible to directly use the intelligent agent prototype to test the interaction and even the algorithm itself. For example, Mercado et al. (2016) compared a HAC system designed using an intelligent agent transparency model with a system designed without using the model, finding that using the intelligent agent transparency model improved team performance and trust levels. For the new interfaces and interaction learning mentioned above, cognitive computation models can help to implement quantitative tests in the early stages of system development (Foyle & Hooey, 2007). When promoting the system to complex application scenarios, scalable and ecological methods or natural scene research (in the wild, Rogers & Marshall, 2017) can be used, the latter is often applied to some non-life-critical systems. Due to the learning of intelligent agents leading to the iteration of HAC systems, longitudinal studies that track team performance over a long period have become one of the best solutions for studying team relationships.

3.2. Scenarios and Platforms

Most contemporary research on HAC is rooted in military contexts, covering command and control tasks such as target identification, attack and defense strategies, navigation, and more. A growing body of HAC research extends to non-military arenas, including agriculture, space exploration, and factory production (O'Neill et al., 2022). These research outcomes are often implemented in intelligent air traffic management systems, intelligent robotic teams, advanced flight deck systems, and autonomous vehicles. At the same time, the focus of HAC research is increasingly shifting toward more general and complex scenarios. For instance, researchers now incorporate higher levels of uncertainty, simulate emergency events, or introduce fuzzy random noise (Xu, 2022), aiming to explore how to maintain a dynamic equilibrium of human-intelligence team performance in rapidly changing environments.

Given the complexity of real-world HAC tasks, most studies employ simulation-based approaches. Various simulation testbeds have been developed to investigate human-intelligence collaboration across different tasks. Military-oriented platforms are especially prominent. Two notable examples include the Comprehensive Task Environment for Unmanned Aerial Systems-Cognitive Engineering Research on Team Tasks-Unmanned Aerial System-Synthetic Task Environment (CERTT-UAS-STE) (Cooke & Shope, 2004) and the Mixed-Initiative Experimental Testbed (MIX testbed, Barber et al., 2008). CERTT-UAS-STE is modeled on the U.S. Air Force's unmanned aerial vehicle ground control station, where the roles of pilot, navigator, and photographer collaborate to photograph specific waypoints. Team members rely on prompt communication (McNeese et al., 2018), and all three roles can configure the AI's capabilities as needed. The MIX testbed comprises an operator control unit and an unmanned vehicle simulator that supports varying levels of automation for reconnaissance, surveillance, and target identification. In this system, intelligent agents can manage vehicle spacing, propose route modifications, or autonomously execute functions once authorized, thereby facilitating active coordination with human operators.

In non-military settings, a typical simulation environment is the Blocks World for Teams, in which humans and intelligent agents jointly move colored blocks from one room to another (Harbers et al., 2011). Researchers have also designed more realistic human-machine resource allocation platforms (computer-human allocation of resources testbed), requiring subjects and intelligent systems to jointly complete the task of preventing resources (such as police preventing crime, Bobko et al., 2022) or area exploration (e.g., Yuan et el., 2022) within a limited time. These and similar testbeds enable controlled experimentation, allowing researchers to translate design and operational challenges into measurable variables.

3.3. Common Research Topics

In the history of human-team-related research, the Input-Mediator-Output (IMO) model has been used to understand teams, organize the teamwork literature, and unpack team interaction and process (O'Neil et al., 2023). Several studies adopting the systematic review approach have summarized current research topics within the HAC research domain (Berretta et al., 2023; Committee et al., 2022; Vaccaro et al., 2024; O'Neil et al., 2023), and these are synthesized using the IMO model in Figure 2.

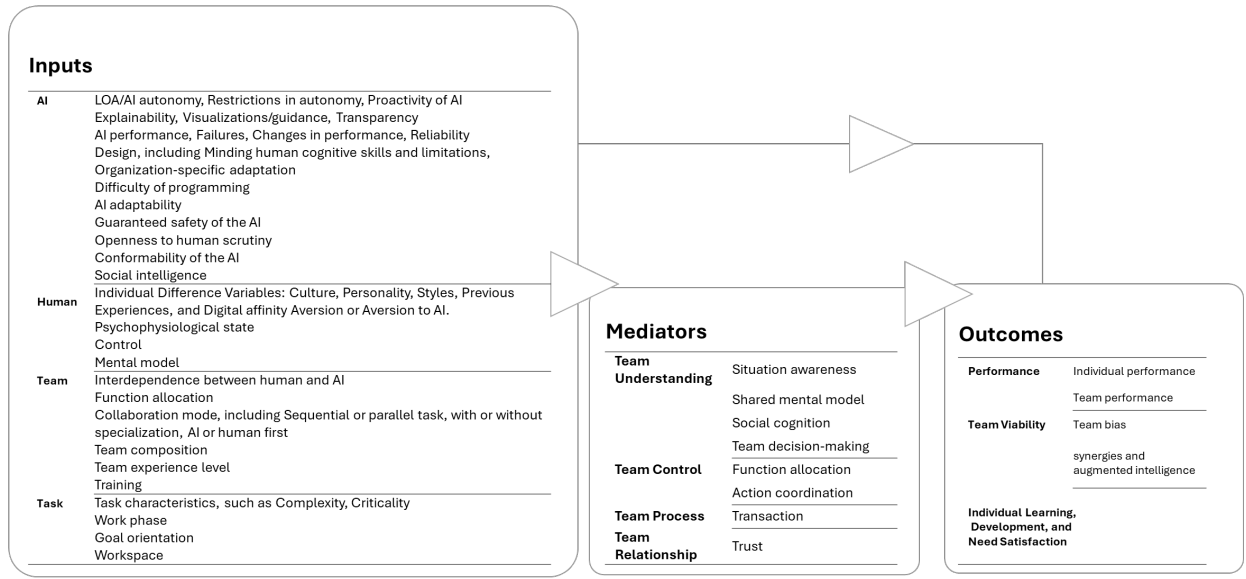

Figure 2 the Input-Mediator-Outcome model integrating multiple frameworks (Berretta et al., 2023; Committee et al., 2022; Vaccaro et al., 2024; O'Neil et al., 2023;). Inputs are theorized to influence mediating mechanisms, which in turn affect multi-level HAC.

(1) Inputs

The inputs to HAC can be categorized into four main areas: AI-related factors, human-related factors, team-related factors, and task-related factors.

AI-related factors such as the LOA and reliability play a crucial role in determining the effectiveness of human-AI collaboration. High LOA intelligent agents often positively impact user attitudes and overall team performance, though this relationship may sometimes follow an inverted U-shape where moderate LOA is optimal (Biondi et al., 2019). Reliability, defined as the accuracy of the information provided by the system, is essential for fostering trust between team members. Systems capable of dynamically adjusting their reliability based on team performance can avoid issues like over-trust caused by complete reliability (Rodriguez et al., 2023). Transparency in information exchange is another key factor. It enables team members to infer each other's intentions more effectively (Kridalukmana et al., 2020), improving trust calibration and overall team performance (Bobko et al., 2022). Moreover, the presentation and content of information can significantly influence communication. Systems that clarify their beliefs and goals (van den Bosch et al., 2019) or utilize social cues, such as combining language and nonverbal communication (de Melo et al., 2021), enhance the quality of interaction. When errors occur, explanations and apologies can mitigate their impact and improve subsequent team performance (Kox et al., 2021).

Human-related factors include individual differences in cognition, personality traits, and experience with intelligent systems. Users with prior experience tend to trust intelligent agents more, are better equipped to handle complex multi-task scenarios (Chen et al., 2011), and are less prone to over-trust during interactions (William et al., 2016). Anthropomorphic systems, which mimic human-like behaviors or use dialogue-based prompts, are more likely to gain user trust (de Visser et al., 2016; McTear et al., 2016). Similarly, intelligent agents that share task goals with users tend to inspire higher levels of trust (Verberne et al., 2012).

Team-related factors encompass the composition, experience level, and the nature of interdependence among team members (Berretta et al., 2023). The composition of the team, including the balance between human and AI participants, significantly influences the team's capabilities and dynamics. Experienced teams tend to adapt more effectively to changes in task demands and can anticipate the behavior of AI agents, enhancing overall performance. The interdependence between humans and AI can be sequential, where tasks are handed off between humans and AI agents, or parallel, where both parties work simultaneously on different aspects of the task. Teams that optimize this interdependence through role assignment and shared understanding tend to perform better.

Task-related factors such as task complexity and interdependence further shape human-AI collaboration. Tasks requiring close human-machine cooperation foster interdependence, which reduces individual workload (Walliser et al., 2017). However, increased task complexity can negatively affect team performance, potentially leading to errors or inefficiencies. Models like Kridalukman et al.'s (2020) supportive SA framework help mitigate this by dynamically adjusting information presentation and prioritizing tasks based on difficulty, ensuring that users maintain an optimal level of SA.

(2) Mediators

Several mediators influence how inputs translate into outcomes in HAC. These mediators encompass team cognition, team control, team transaction, and team relationship. The first three elements parallel the human team processes defined by Marks et al. (2001): transition, action, and interpersonal activities.

Transition-related activities involve developing, adapting, and clarifying the team's common purpose, strategy, tasks, and role structure. These activities ensure alignment between humans and AI agents regarding goals and roles within the team. In HAC contexts, team cognition serves as the foundation for planning dynamic interactions (Cooke et al., 2013), with prerequisite development of shared mental models and SA, both contributing to final decision-making outcomes (Endsley, 2017). Effective information sharing enables team members to align goals and anticipate each other's actions. For example, dynamic sharing of task-relevant information enhances the team's SA and fosters a more cohesive understanding of the situation.

Action-related activities occur during task execution and involve role assignment based on planning, backup behaviors, mutual performance monitoring, tracking goal progression, and coordination. In HAC, team control typically begins with function allocation followed by action coordination during task execution. Function allocation can be understood through role theory, where responsibilities are distributed among team members based on their designated roles (Crawford & Lepine, 2013). According to these allocated functions, autonomous agents may provide real-time feedback on task status or adjust their actions to coordinate with humans, offering support during critical moments.

Both transition and action processes depend on interpersonal activities, where affect management and conflict resolution occur. In HAC, while these cannot be strictly classified as "interpersonal" since AI is not a person, transactions between humans and AI occur whenever team cognition and team control are present. Communication, both verbal and non-verbal, serves as the primary channel for these transactions.

Finally, team relationship mediates the interaction between inputs and outcomes by shaping HAC dynamics. Trust remains the cornerstone of effective team relationships (Makovi et al., 2023). Proper trust calibration ensures that humans neither over-rely on nor underutilize AI capabilities, which is particularly crucial in life-critical domains such as healthcare or transportation (Sagona et al., 2024).

(3) Outcomes

The outcomes of HAT can be observed across three dimensions: performance, team viability, and individual learning and satisfaction. Performance is measured through both individual and team perspectives. For example, the time required for human operators to take over control during emergencies reflects the effectiveness of collaboration (Demir et al., 2019; McNeese et al., 2018). Task efficiency, completion time, and the quality of overall planning are key metrics for evaluating team performance (O'Neill et al., 2022). Team viability reflects the long-term stability and adaptability of the team. Factors such as trust calibration, the mitigation of biases, and the synergy between humans and AI contribute to the sustainability of the team's operations. These aspects ensure that the team remains functional and effective even under changing circumstances or task demands. Individual learning and satisfaction are also critical outcomes of HAC. Through effective collaboration and task sharing, users gain valuable experience, develop new skills, and achieve a higher level of satisfaction with their role in the team. These factors not only enhance individual performance but also contribute to the overall success of the human-intelligence partnership.

## 4. Current Research Agenda of HAC

The success of collaboration extends beyond team inputs such as member capabilities and available resources, hinging critically on the mediators of collaboration within the team (see Mediators in 3.3 and Figure 2). These mediators play a fundamental role in determining collaborative outcomes. The following sections will examine each of these mediators in detail to provide a comprehensive understanding of HAC research.

4.1 Team Cognition

Team cognition is a key content in HAC research, which refers to cognitive activities that occur at the team level and usually form a common understanding within the team. Team cognition is not a simple addition of individual knowledge, but a shared knowledge system that develops over time and context through interaction among team members (Cooke et al., 2013). The change of machine roles in HAC determines that the smooth cooperation of human-machine teams requires efficient team cognition interaction sharing modes (Xu, 2022). Existing research has found that both cognitive and social cognitive factors, such as shared mental model (SMM), team situation awareness (TSA), social cognition, and team decision will affect the team performance and coherence of HAC systems.

(1) Mental Model

Mental model is an internal cognitive representation of the external real world influenced by factors such as experience, beliefs, and rules. It is a cognitive structure that people use to describe task objectives, explain system functions, observe environmental states, and predict future situations (Morris & Rouse, 1985). As an important cognitive mechanism for organizing information and assisting decision-making, the mental model can guide individuals or groups to take specific actions and measures based on the description, understanding, and prediction of the internal and external environment. It will affect the processes of attention allocation, reasoning interpretation, and behavior selection in individual actions and team cooperation.

In team cooperation contexts, members must construct a SMM, forming a coordinated and unified cognitive framework at the team level. SMMs represent knowledge structures shared among team members (encompassing environment, goals, plans, and behavioral intentions), enabling accurate representations and predictions in dynamic task situations while facilitating synchronized understanding of situations and actions. In team tasks, SMMs can be categorized into task mental models (specific knowledge structures for understanding and completing common tasks) and team mental models (knowledge for understanding the team and teammates' beliefs and characteristics), each influencing team cooperation cognitive processes from different perspectives (Andrews et al., 2022). The importance of forming team mental models is heightened by the distinct behavioral patterns exhibited by AI agents, which differ significantly from human behaviors. This divergence introduces additional complexity into human judgment and comprehension of AI agents within HAC, making such interactions more intricate than those in human-human teaming (Miller, 2019). Furthermore, as AI agents develop greater autonomy with independent behavioral capabilities that introduce output uncertainty, their operational processes become increasingly opaque and difficult for human operators to comprehend (Casner & Hutchins, 2019).

HAC imposes higher requirements for SMM formation, with human-centered approaches emphasizing active alignment of AI to human values. Value alignment has philosophical foundations regarding machines pursuing unintended objectives, articulated as "the purpose put into the machine is the purpose which we really desire" (Nobert, 1960). Value alignment constitutes a prerequisite for establishing meaningful relationships between human operators and artificial agents, as humans can more readily predict and interpret the behavior of AI systems whose values mirror their own (Gabriel, 2020). Recent AI researchers have advanced the value alignment problem through methods such as Cooperative Inverse Reinforcement Learning, which conceptualizes AI-human interaction as a cooperative process where AI systems learn human objectives through observation and gradually align their operations accordingly (Russell, 2019; Yuan et al., 2022). Additionally, Microsoft's Value Compass initiative represents an institutional approach to aligning AI systems with human values through the integration of sociological and ethical perspectives, with the objective of guiding AI behavior in accordance with societal norms and expectations (Yao et al., 2025). Collectively, these multidisciplinary approaches represent endeavors in the HAC field to embed human values within AI architectures, ensuring that advancements in AI capabilities proceed in consonance with human intentions and ethical frameworks.

(2) Situation Awareness

In team cognition research, how team members maintain SA, how they share SA among themselves, and what factors affect the establishment and loss of TSA are all issues of concern to researchers. The concept of SA was first proposed by Endsley (1988) to reflect an individual's perception, comprehension, and projection of dynamic environmental elements (including system environment, task clues, team cooperation, interaction interface, etc.) in time and space. This model divides SA into three levels: the perception level, where relevant environmental elements are perceived; the comprehension level, where the perceived elements are processed and understood based on goals and experience; and the projection level, where future conditions are predicted based on the analysis and understanding of the information. SA is built on top of mental models, essentially by extracting relevant information from mental models in long-term memory into working memory, where cognitive processing happens (Endsley, 1988).

When SA is placed in the context of team cognition interaction, it forms TSA, built on top of the team SMM, together constituting a complete HAC cognitive architecture. TSA is the ability for all members in a team collaboration scenario to acquire SA related to their tasks and responsibilities, aiding the team in making collective decisions (Endsley, 2020). TSA allows team members to obtain appropriate information from the right sources at the right time, enabling the entire team to have synchronous perception, understanding, and prediction of the task and environmental conditions, ensuring coordinated decision-making (Gorman et al., 2017). In high-level team collaboration, the team needs to understand specific situational characteristics (such as clues, conditions, and patterns in the environment) and factors related to the task status (such as goals, decisions, and function allocation within the team) to establish TSA. Good TSA relies on the sharing of SA within the team and requires appropriate team communication and interaction methods (Demir et al., 2019).

Research on TSA has traditionally been confined to human-human collaboration, as non-intelligent machines inherently lack machine SA, preventing the establishment of mixed human-machine TSA. However, the emergence of HAC has fundamentally transformed this paradigm, with AI agents now functioning as autonomous teammates capable of independent decision-making and problem-solving. Models incorporating AI agents' SA have demonstrated effectiveness in explaining and predicting HAC (Salmon & Plant, 2022), and research into human-AI

SA sharing has become essential for building human-AI TSA and maintaining interactivity (bidirectional synchronous situation perception), collaboration (task cooperation and decision sharing), and communicability (efficient information exchange) in HAC systems. Consequently, AI agents must develop cognitive architectures corresponding to human psychological processes to facilitate effective information exchange with human counterparts, enabling compatible SA to be shared across both human and AI team members. This aligns with calls from the research community to construct bidirectional interactive cognitive systems within HAC (Xu et al., 2024).

Based on this analysis, Gao et al. (2023) integrated shared cognition and interactionism perspectives to propose the Agent Teaming Situation Awareness (ATSA) model for HAC (Figure 3). This model unifies team cognition, team control, and team process through the lens of TSA formation in HAC contexts. Centering on SA and mental models from team cognition, ATSA conceptualizes humans and AI as autonomous intelligent agents achieving team goals through both individual and team-level interactions. At the individual level, drawing from perceptual cycle theory (Neisser, 1976; Smith & Hancock, 1995), the model delineates a comprehensive SA process for each intelligent agent, where mental models direct external behavior, allowing agents to sample and modify their external environment through actions. Subsequently, environmental changes are reflected in the evolving mental models of the agents. The team layer describes TSA, encompassing teaming understanding, teaming control, and the world. Teaming understanding and teaming control extend the mental models and actions defined at the individual level. The interaction between individual and team layers operates bidirectionally, facilitated by transactive processes—primarily achieved through team communication. This bidirectional flow enables dynamic influence between layers, enhancing overall team performance. The world serves as the connection between humans and AI agents, with both individual and team layers interacting with and modifying the same environment.

As a cognitive model, ATSA's core components are rooted in individual mental models and team shared understanding. It integrates SA (or TSA) with mental models (or SMM) by conceptualizing SA as the activated subset of these cognitive structures. The model posits that SA emerges from activated components of the mental model—those with higher informational priority and greater accessibility compared to non-activated information. The ATSA framework employs a principle of heterogeneous homomorphism between human and AI agents: AI agents possess more flexible functional capacities—represented by dashed lines—than their human counterparts. Accordingly, the model advocates for a human-centered approach through adaptive and adaptable AI systems (Calhoun et al., 2022), a concept termed "plastic AI empowering humans." At the behavioral level, human-centeredness is maintained through ultimate human control. Both humans and AI agents may, under certain circumstances, exert excessive control that jeopardizes task performance or undermines fundamental human values such as safety and dignity. Thus, effective team control mitigates negative outcomes of such overreach through control locks or compensatory mechanisms, thereby preserving the integrity of HAC.

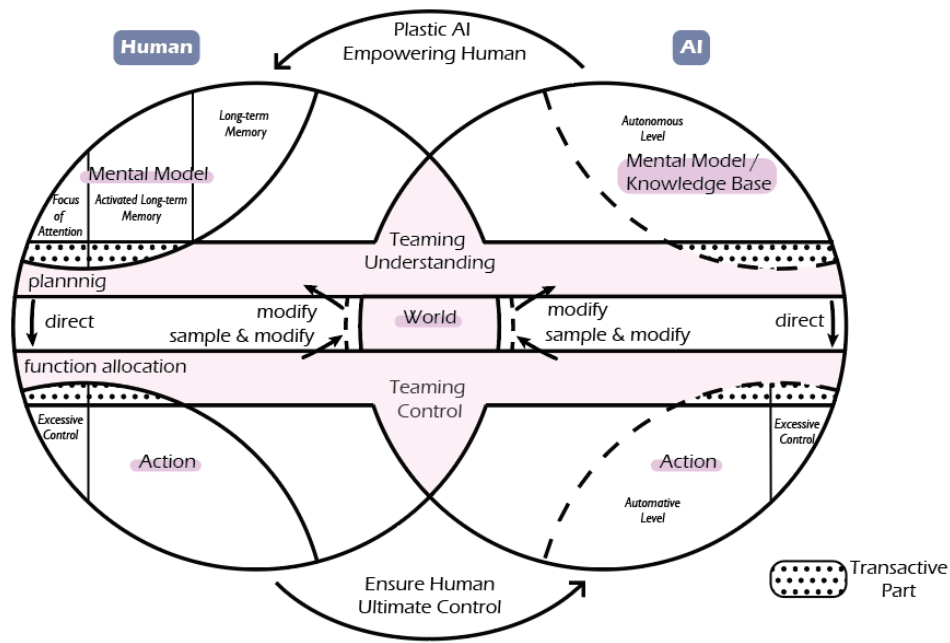

Figure 3 ATSA Framework

Existing research has explored the construction methods and influencing factors of TSA in HAC situations from the perspective of team interaction. Demir et al. (2017) found that team members' expectations of teammates' (human or intelligent agent) information needs are important factors in building TSA. HAC needs to improve the information push mechanism between team members, anticipate teammates' situational information needs in verbal communication, and proactively push information related to their task goals, making the construction of human-machine TSA more efficient (Cohen et al., 2021). Demir et al. (2019) suggested that establishing an efficient human-machine communication coordination mechanism is very necessary for maintaining TSA in the HAC system. Realizing the sharing and exchange of SA between humans and intelligent agents through a multimodal cooperative human-machine interaction interface can help improve TSA (Endsley, 2020). In addition, transparency and explainability play important roles in the process of building TSA in HAC, which will be elaborated in the team process section.

(3) Social Cognition

Social cognition is the process of encoding and decoding various behaviors in social interactions, it plays a crucial role in shaping human interactions and behaviors, as well as understanding self, others, and interpersonal relationships. In HAC, users tend to interact with AI agents in a social way, therefore, effective social cognition relies on endowing AI with social intelligence, which refers to the ability to interact effectively with others to achieve goals (Ford & Tisak, 1983). Contemporary research focuses extensively on constructing artificial social intelligence that enables AI with enhanced social awareness and capabilities to facilitate more natural interactions (Fan et al., 2022; Bendell et al., 2025). Consequently, AI agents that can perceive stimuli from the outside world like humans, understand, process, and make emotional responses are rapidly emerging. Current AI systems demonstrate significant proficiency in multimodal emotion recognition and basic theory of mind (Freeman et al., 2022), but require substantial improvement in higher-order theory of mind and cognition of complex social environments (Wang et al., 2024).

While social intelligence requires AI systems to demonstrate social awareness and social competencies, humans need to develop new ways of establishing rapport with AI teammates (Duan et al., 2024). Through deliberately designed appropriate social traits (e.g., appearance, social roles, social competencies) and behaviors (both verbal and non-verbal), AI agents can modulate human social cognition, fostering appropriate team experiences and trust relationships (Li et al., 2023; Kim et al., 2023). For instance, as a strong social signal, gaze can modulate human's behaviors, strategies and neuroactivities (Belkaid et al., 2021).

AI systems with advanced social skills are better able to interact with humans, leading to higher team performance in complex contextual interactions with humans, particularly in socially nuanced environments. In HAC, social AI systems can achieve more natural interaction through non-verbal communication methods, especially emotional communication, to communicate and cooperate with humans, establish trust, and reduce errors (de Melo et al., 2021; Yasuhara & Takehara, 2023; Lavit Nicora et al., 2024).

(4) Team Decision Making

The cognitive process of a team revolves around team decision-making (O'Neill et al., 2016). The team decision-making in HAC emphasizes the shared decision-making power between humans and intelligent agents. An effective HAC should allow the sharing of decision-making power between humans and intelligent agents at various levels such as tasks, functions, and systems (Xu, 2022). When HAC realizes the sharing of decision-making power between humans and machines, it needs to reasonably allocate decision-making power based on the results of information integration processing. At the same time, members of the human-intelligence team should provide real-time two-way information interaction feedback, and continuously adjust and improve the decision-sharing mechanism according to the actual situation. The "human-centered" HAC relationship emphasizes that the role of AI is to enhance human capabilities rather than replace humans, and humans should have the final decision-making authority of HAC. Therefore, HAC needs to establish a complete and reliable mechanism for the allocation and transfer of human-machine decision-making power. In the daily operation of the team, the sharing of human-machine decision-making authority promotes the efficiency of team decision-making, while ensuring that humans have the final decision-making power and can execute smoothly at decisive or critical moments. Currently certain explorations on team decision-making in HAC has been conducted, including factors affecting team decision-making in HAC, ways of allocating and transferring team decision-making authority, etc.

4.2 Team Control

The current research on the process layer of HAC mainly revolves around function allocation and action coordination. To avoid failure, errors, and inoperability, HAC must support real-time dynamic function allocation. At the same time, action coordination is the main path for achieving the allocated task in HAC.

(1) Function Allocation

Function allocation refers to the strategy of allocating system functions and tasks between humans and intelligent agents. HAC emphasizes dynamic allocation (Kaber, 2018): neither humans nor intelligent agents have fixed responsibilities, and they need to flexibly determine their respective tasks to achieve their goals. This dynamic allocation requires the system to maximize human-intelligence complementarity on the premise of being human-centered. Therefore, humans should have as much control as possible in the process of task allocation; while machines can adaptively adjust the LOA according to the advantages and disadvantages of humans in the corresponding situation. However, some researchers suggest that since humans and intelligent agents are collaborative partners in HAC working toward shared goals, the task allocation authority between humans and machines also needs to be more equally shared (Johnson & Vera, 2019). This type of human-intelligence function allocation research mainly draws on the research ideas in human-human teams, allowing humans and intelligent agents to communicate task goals and plans, and choose the LOA and task allocation between human operators and intelligent agents before or during operation (Miller & Parasuraman, 2007).

Dynamic function allocation means that human-machine teams need to find the most suitable function allocation method for the current task situation in the operation process to maximize team efficiency (Kaber, 2018). When the LOA of the intelligent agent exceeds the appropriate range, it needs to be readjusted to the proper level (Miller & Parasuraman, 2007). This adjustment reflects the transfer of control (TOC), the active or passive adjustment of the system LOA triggered by HAC in a certain situation (Lu & de Winter, 2015). Currently the widely accepted and applied TOC classification framework in the field mainly considers two key factors: the initiator (human-initiated vs. AI-initiated) and the direction of conversion (human control to AI control vs. AI control to human control), dividing the main ways of human-machine control transfer into four categories through the combination of these two factors (McCall et al., 2019). Among them, takeover represents the process of transferring control to humans initiated by AI, which guarantees that humans have the final decision-making authority in HAC. When the intelligent agent encounters situations that it cannot cope with or specific emergencies, the intelligent agent needs to quickly transfer control of the relevant task to humans and provide corresponding situational information. The takeover issue in life-critical systems like autonomous driving is a current research hotspot. Handover refers to the TOC process of transferring system control to intelligent agents initiated by humans, usually as a non-emergency TOC method (Miller et al., 2014). Handovers have received relatively less attention in research, as it is a straightforward process and typically occurs in non-emergency situations. However, due to its unique role in promoting the interdependence of human intelligence and enhancing human expectations of the system's capabilities, it remains an indispensable component of human-AI system TOC. The other two TOC methods, where team members directly request to obtain system control, are called override, and the initiator and receiver of this control transfer process are the same. The override where the intelligent agent as the initiator transfers system control from humans to intelligent agents is called intelligent agent override, which rarely occurs only when the human operator performs too poorly. The override where humans as the initiator transfer system control from intelligent agents to humans is called human override; although it is a widely used TOC method in many situations, current research is relatively less. Subsequent research added factors such as time, compulsion, and predictability based on this classification (Lu et al., 2016).

Adaptive LOA adjustments, compared to fixed LOA, can improve task performance to a certain extent and reduce the demand for human attention (Parasuraman et al., 2009). However, when the system automatically changes the LOA, human acceptance of automation may decrease, and humans may lack SA to complete the corresponding operations. Especially in complex human-machine cooperative task situations, incorrect, untimely, and unpredictable transfers of LOA by adaptive systems may become annoying factors that even affect system performance and safety. In contrast, the adaptable mode allows humans to make decisions about the system's adaptive adjustments to LOA, better ensuring the "human in the loop" in human-machine cooperation. Adaptable systems emphasize preserving human authority to change LOA, with the AI agent suggesting several options, and the LOA changing process is carried out according to human choice. However, the adaptable mode cannot necessarily ensure that the system is at the most suitable LOA, and humans may choose an inappropriate LOA due to being engrossed in irrelevant tasks or being careless. At the same time, compared to adaptive systems, adaptable systems require humans to undertake additional decision-making tasks, thereby increasing human workload. It is worth noting that Calhoun (2022) summarized comparative experiments between adaptive and adaptable automation, suggesting that humans tend to prefer adaptable automation. It not only offers advantages in improving system task performance but also helps reduce the perceived workload for humans.

(2) Action Coordination

There are two types of control in HAC, shared control and cooperative control (Marcano, 2020). Cooperative control refers to human and AI are responsible for their own subtasks, together constates a global task, while shared control refers to that human and AI agent need to conduct continuously on the same task. Compared with cooperative control, shared control introduces the problem of action coordination, where humans and AI synchronize on actions for achieving a specific task unit. This, of course relies on the action transaction within the team, through exhibited behaviors and overt communications (e.g. Angleraud et al., 2021). However, it also can be achieved through non-verbal cues. These kinds of implicit coordination, defined as the ability to act in concert by predicting needs and adjusting behavior without overt communication, is crucial for action coordination in dynamic settings, where explicit communication might be impractical by (Rico et al., 2008).

The interaction modes reflect the ways in which humans and AI agents coordinate their actions, relying on a series of complex dynamic processes that include, but are not limited to, information exchange, task collaboration, decision support, and problem-solving. Gomez et al. (2025) outlined seven distinct interaction modes between humans and AI, with the delegation mode referring to a complete transfer of decision-making and control from humans to AI. The remaining six modes involve cooperative human-AI interactions, which the author further categorized based on information processing and interaction initiation. These six modes are as follows: AI-first, Secondary, AI-guided, AI-follow, Request-driven, and Human-guided.

Table 6 HAC Interaction Modes

| Action Initiator | Information Explorer | | |
|---|---|---|---|
| | **Parallel** | **AI Agent** | **Human** |
| **AI Agent** | AI-first | Secondary | AI-guided |
| **Human** | AI-follow | Request | Human-guided |

1. **AI-first**: In this mode, human-AI collaboration involves a series of decision-making processes. The AI presents both information relevant to the decision and its predicted outcomes, which the human can either consider or disregard before arriving at their final decision.

2. **Secondary**: The AI provides supplementary decision-making information without offering a decision outcome. The human uses the information provided by the AI to aid in their decision-making process.

3. **AI-guided**: In this mode, the AI requests information from the human. Guided by the AI's instructions, the human responds by providing the necessary information. This iterative exchange continues until the AI has sufficient data to make a prediction. The AI then presents its decision prediction, and the human uses this prediction to make their final decision.

4. **AI-follow**: The human develops an initial independent prediction for the given decision task. The AI then presents its predicted outcomes along with related information, which the human integrates with their own prediction to make the final decision.

5. **Request-driven**: The AI does not automatically generate predictions or provide supplementary decision information. Instead, it only responds when the human actively requests predictions or decision-support information. The human, based on their needs, prompts the AI to provide relevant predictions or information and then makes the final decision independently or in combination with the AI's input.

6. **Human-guided**: The AI provides its predicted outcomes and relevant supporting information. The human can correct the AI's erroneous decisions, demonstrate how to make the correct decisions, or guide the AI to make more optimal decisions. This iterative process of information exchange continues, leading to progressively improved decisions from the AI until a satisfactory outcome is achieved.

4.3 Team Transaction

Both team cognition and team control are achieved through the transaction among individual cognition and control, and these transactions are mainly various kinds of communication. Communication is a reciprocal team process in which two or more team members exchange information through verbal or non-verbal channels. It ensures the formation of shared SA, mental models, and team goal consistency (Lyons et al., 2021), and can provide team members with emotional and social cues to improve team efficiency, helping humans to perceive machines as teammates (Iqbal & Riek, 2017). Efficient teams need to collectively perceive the environment, understand the relationship between the environment and common goals, and take communication processes based on these goals to support others in the team.

Current human-intelligent agent communication primarily occurs through visual representations, text-based exchanges, and audio expressions (O'Neill et al., 2022). These modalities may be employed independently or in combination. Similar to human-human teams, which utilize behavioral cues for implicit communication such as directional actions and gaze signals, communication in HAC must maintain consistency between verbal statements and actions (Banerjee et al., 2018) and dynamically transmit information through contextually appropriate methods (Bengler et al., 2020). Recent research indicates that proactive communication from AI systems (such as providing updates and clarifying intentions) significantly enhances coordination, whereas AI systems employing other communication strategies—including immediate responses, balancing efficiency with sociability, and adhering to established communication patterns—are less likely to be perceived as effective teammates (Zhang et al., 2024).

Although communication is inherently bidirectional, advocating for AI systems to communicate effectively with human partners remains a critical concern in HAC research, often discussed in terms of explainability and transparency. Explainability conveys information about the reasoning processes and decision-making logic within the "black box" of intelligent systems to human collaborators, thereby enhancing Team Situation Awareness (Endsley, 2023), calibrating trust in AI capabilities (Schemmer et al., 2023), and improving overall task performance (Senoner et al., 2024). However, explanations from AI systems prove beneficial only when delivered with appropriate interaction frequency, content relevance, communication modality, and timing (Demir et al., 2019; Lyons et al., 2021; see meta-analysis in Vacarro et al., 2024). As Miller (2019) observed, "explanations are not just the presentation of associations and causes (causal attribution), they are contextual." Human users require explanations that are contrastive (addressing counterfactual scenarios), selective (focusing on a limited number of causal factors), and social (aligned with beliefs, desires, and intentions). Furthermore, users prefer "practically useful" explanations that enhance collaboration rather than technical details or probability assessments (Kim et al., 2023; Miller, 2019).

Explainability can be achieved through high transparency, which aims to inform users about the behavior, reliability, and intentions of intelligent agents, enabling people to understand their reasoning processes, task performance capabilities, intentions, and future plans (Bhaskara et al., 2020). Two widely recognized models of HAC transparency were proposed by Lyons (2013) and Chen et al. (2018). The former suggests that AI agents should be transparent to humans regarding intention, tasks, analysis structure, and environment, while humans should be transparent to agents regarding team tasks and human states. The latter introduces a transparency model based on three levels of Situation Awareness.

However, transparency is not a prerequisite for AI explainability. When team cognition is consistent and members share common task and team mental models, they can form accurate expectations of teammates' intentions through team cognition (Chen et al., 2018). Therefore, explicit communication is not always essential for effective team collaboration. Similarly, a machine's effectiveness is maximized only when closely aligned with appropriate human expectations. Moreover, excessive transparency may disrupt social interaction cues. For instance, when an AI agent demonstrates only lower-order capabilities, people may assume these represent the agent's full potential and abandon efforts to collaborate on more complex tasks (Fischer, 2018). Consequently, achieving interpretability of intelligent agents remains a significant challenge in the HAC field, requiring interdisciplinary collaboration across algorithm development, psychology, human factors, sociology, and related disciplines.

4.4 Team Relationship: Trust

Relationship fulfillment represents one of three fundamental human psychological needs (Ryan & Deci, 2000), and assumes particular significance when examining interactions with AI systems that exhibit human-like social characteristics. Within this context, trust functions as a critical factor as humans evaluate AI capabilities and reliability, encompassing both ability trust regarding task performance and integrity trust concerning adherence to behavioral principles (Hussain et al., 2021). Human trust in artificial systems first garnered attention during the automation era and has since evolved into an essential consideration in the current intelligence era. A widely accepted conceptualization of human-machine trust comes from Lee and See (2004), who define it as an individual's attitude wherein they believe that machines can facilitate the achievement of specific objectives under conditions of uncertainty or vulnerability. This trust framework provides a foundation for understanding the complex relational dynamics that emerge between humans and AI collaborators in professional contexts.

Taking autonomous driving as an example, Gao et al. (Gao et al., 2021; Chen et al., 2021) analyzed the development process and influencing factors of trust in the HAC process, and proposed a dynamic trust framework based on the trust development process (Figure 4). This framework divides the development of trust into four stages: dispositional trust, initial trust, ongoing trust, and post-task trust. At different stages of trust development, the factors affecting trust are different. These factors can be summarized as operator characteristics, system characteristics, and situational characteristics. Among them, operator characteristics can be divided into two factors: inherent traits and prior knowledge. Inherent traits refer to the relatively stable characteristics of an individual that are physiologically inherent or formed over a long period, such as gender, personality, age, cultural background, etc., which are unrelated to the system and situation; prior knowledge refers to the system and situational characteristics obtained through learning. In the interaction process between the operator and the intelligent system, the system characteristics and situational characteristics are objectively reflected through the system performance. This objective performance is then processed by the operator's cognitive system into subjective perceptions, including the individual's perception of potential risks associated with the system's performance. These subjective perceptions are direct factors that influence trust. Among them, ongoing trust and its influencing factors are the focus of current research. Appropriate ongoing trust depends on the operator's accurate perception of the system and situational characteristics (the operator's SA level). As shown in Figure 4 with an arrow circle, real-time trust affects the operator's dependent use behavior of the system. In the interaction process, operators can dynamically perceive the system characteristics and situational characteristics through the system performance, and adjust their real-time trust level dynamically based on this.

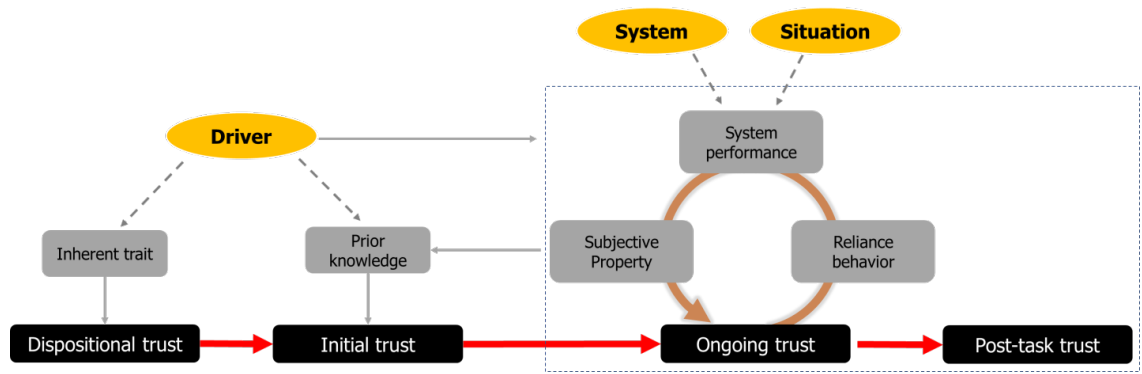

Figure 4 Dynamic trust framework based on the trust development process. The upper line from "driver" to the blue-dashed line box represents that the operator's user characteristics affect all other four factors except system performance, and the lower line from the box to "prior knowledge" represents that all factors in the box can be converted into the user's prior experience.

The purpose of studying trust is to calibrate trust, maintaining an appropriate level of trust in machines by human operators. Not surprisingly, studies on human-machine trust in HAC mainly revolve around avoiding over-trust, avoiding insufficient trust, and trust repair in HAC. Avoiding insufficient trust aims to prevent trust crises through design means before they occur, while trust repair is remedial after trust is damaged. For trust calibration, researchers currently mainly consider from the aspects of monitoring correction, operator training, and optimizing human-machine interface design (e.g., Gao et al., 2024; Kox et al., 2021; William et al., 2016).

## 5. Human-Centered Human-AI Collaboration

To adequately address the research agenda previously outlined, HAC studies require a fundamental shift in the perspective, from technical problems to human-centered AI that prioritizes *human involvement and input* in decision-making processes and ensures *human leadership and control* within human-AI teams (Garibay et al., 2023; Shneiderman, 2022; Stanford HAI, 2022; Xu & Gao, 2024). Many current studies exhibit a technology-centric bias by emphasizing algorithmic advancement over collaborative dynamics. This leads to the development of systems that, despite technical sophistication, fail to integrate effectively with human cognitive processes and established work patterns, and ensuring meaningful human involvement and input is an essential solution for reducing these cases (Arrieta et al., 2020). Furthermore, non-technology-centric approaches that position humans in equal roles with AI systems are similarly inadvisable. Such positioning undermines human agency and can lead to reduced engagement, trust issues, and suboptimal collaborative outcomes (Bigenwald & Chambon, 2019). As Amershi et al. (2019) observed, systems designed without clear human leadership mechanisms frequently encounter adoption challenges and demonstrate reduced operational effectiveness.

However, in discussions of human leadership within HAC, it is essential to move beyond notions of static leadership where humans maintain exclusive decision-making authority. Instead, leadership in HAC should be conceptualized as dynamic and context-dependent, encompassing varying degrees of human oversight and AI autonomy. For example, in advanced autonomous vehicle systems, AI can manage routine driving tasks while humans retain ultimate authority to intervene or override decisions when necessary (Van den Bosch et al., 2024).

To effectively navigate complex tasks within HAC frameworks, drawing from established leadership theories in human-human teams provides valuable insights. The following sections first review leadership literature on human-human teaming and then develop a conceptual framework for human-led HAC.

5.1 Leadership Insight from Human Teams

Leadership theory in human teams has progressed significantly from trait-based approaches toward more fluid, contextual models (Dinh et al., 2014). In the meantime, leadership frameworks are evolving beyond static models toward dynamic co-evolution of roles, reflecting the complex, adaptive nature of modern organizational environments. This evolution acknowledges that effective leadership does not merely involve fixed positions or unchanging characteristics, but rather encompasses dynamic interactions that respond to evolving circumstances and capabilities (Adriasola et al., 2021).

The traditional static leadership, *vertical leadership,* is characterized by a hierarchical structure where the leader holds authority and is responsible for guiding the team. It describes the top–down leadership of external team leaders, ensuring formalized roles and decision-making processes (Pearce & Conger, 2003). In contrast of this, the *shared leadership* paradigm exemplifies the shift to dynamic perspective, distributing leadership functions across team members based on expertise and situational demands rather than formal authority (Pearce & Conger, 2003; van Knippenberg et al., 2024). What's more, *transformational leadership*, with its emphasis on vision articulation, intellectual stimulation, and individualized consideration (Bass & Riggio, 2006), creates conditions for leadership capacity to develop throughout a team, enabling what DeRue (2011) describes as "leadership as a social process" where leading and following become complementary.

These leadership frameworks offer implications for HAC. An effective leadership structure for human-AI contexts should incorporate principles analogous to shared leadership, termed here as "shared responsibilities". Attributing "leadership" to AI in the same vein as human-human teams is problematic for several reasons. First, AI, in its current and foreseeable forms, lacks the inherent intentionality, ethical judgment, and holistic understanding that underpin human leadership. Second, the concept of "shared leadership" itself is subject to cultural variability; societal norms regarding hierarchy and power distance significantly influence its acceptance and enactment across different cultural contexts (House et al., 2004). Downgrading "leadership" to the more universally understood and accepted concept of "responsibility" allows for a more culturally adaptable framework. Therefore, "shared responsibilities" more accurately and inclusively reflects the collaborative paradigm in HAC, where decision-making and planning accountabilities are dynamically distributed based on the comparative advantages and evolving capabilities of both human and AI actors.

However, this flexibility must be complemented by vertical leadership elements that ensure humans maintain ultimate control and strategic oversight of AI systems. Furthermore, transformational leadership principles provide a framework through which AI can enhance human capabilities rather than diminish human capabilities. Building on these implications, a conceptual framework for human-led AI collaboration will be proposed in the next section, where key factors for achieving human-led and human-centered HAC will be introduced.

5.2 The Conceptual Framework: Human-Centered Human-AI Collaboration (HCHAC)

Two foundational principles emerge in the formation of team control and team cognition process: *human-led ultimate control* and *AI empowering humans*. The first principle underscores the necessity of human oversight in AI operations, ensuring that AI practices remain reliable, safe, and trustworthy (Shneiderman, 2022). The second principle of AI empowering humans focuses on leveraging AI's strengths to augment human's capabilities, thus forming hybrid enhanced intelligence that empowers humans (Xu & Gao, 2024).

Based on the two fundamental principles and three leadership types (vertical leadership, shared leadership (i.e., shared responsibilities in HAC context), and transformational leadership), a framework depicting human-centeredness in HAC is shown as Figure 5. This framework is elaborated based on a simplified perceptual cycle, where both human and AI sample from the shared world, forming mental models, which further guide their actions.

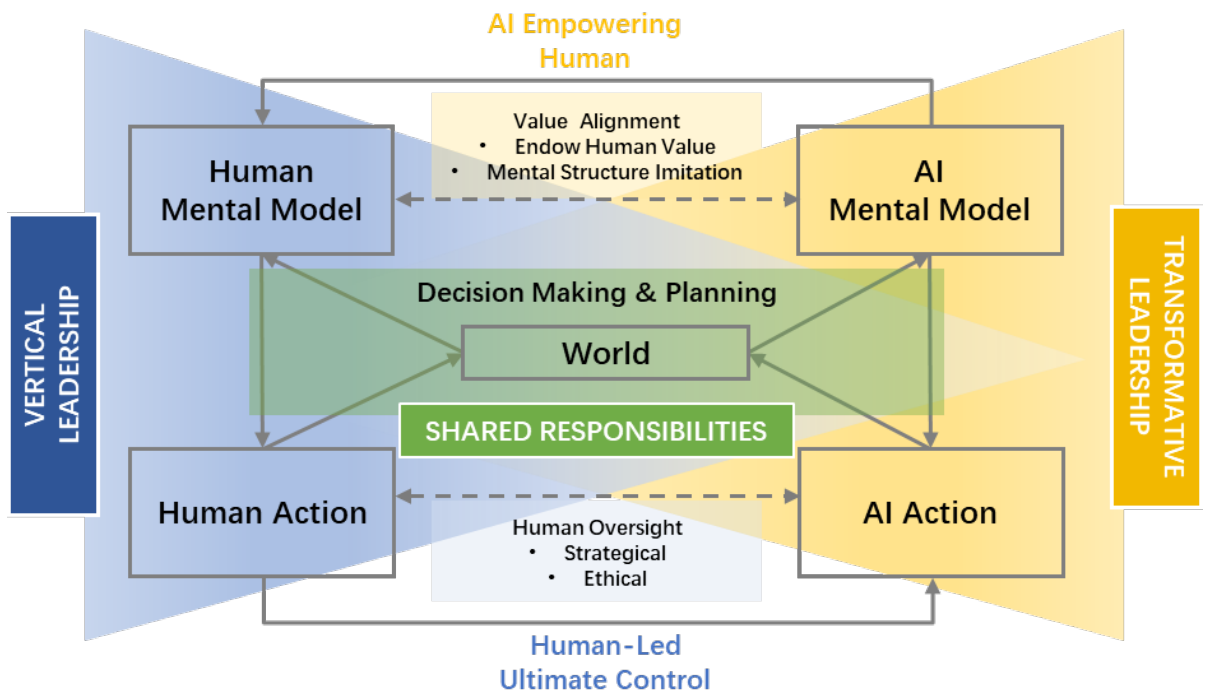

Figure 5 Human Centered Human-AI Collaboration (HCHAC) Framework. The framework delineates two foundational pathways: (1) *AI empowering human* (indicated in yellow) stems from the value alignment of the AI's mental model to human's mental model, enabling transformative leadership from AI toward humans; (2) *Human-led ultimate control* (indicated in blue) originates from human oversight over action-level processes, ensuring that final authority and accountability remain with the human operator, consistent with vertical leadership. These two pathways reflect a bidirectional implementation of human-centeredness, with each originated from human and AI endeavor. The human's vertical leadership secures ethical and strategic governance over AI actions, while the AI's transformative leadership facilitates human augmentation by aligning to the human mental model through value alignment. Notably, *shared responsibilities* (indicated in green) emerge dynamically within the interaction between human and AI, where responsibilities are distributed in real time for decision-making and planning layer.

(1) Principle 1: Human-led Ultimate Control

*Vertical Leadership: Establishing Human Oversight*

The foundation of effective human-AI collaboration should be anchored in human oversight and ethical governance (Floridi et al., 2023). Vertical leadership in HAC establishes clear lines of authority that place humans at the top of the decision hierarchy, particularly for strategic and ethical decisions.

Strategical and ethical oversight encompasses several crucial dimensions including explainability, transparency, bias mitigation, accountability, and fairness. AI systems must provide understandable explanations for their recommendations and actions for human supervision and leadership. With human oversight, identifying, reducing, and addressing algorithmic biases becomes possible (Holstein et al., 2024). Human oversight also ensures AI systems treat all stakeholders equitably and fairly, as shown in a longitudinal study by Garcia and Lee (2024) where regular human audits of AI decision patterns significantly improved fairness metrics across multiple domains. Through the vertical human leadership, clear frameworks for assigning responsibility when AI systems make errors or cause harm also becomes possible, which is of urgent need as organizations with established accountability frameworks experienced fewer adverse incidents and faster remediation (Jobin et al. 2022).

Beyond oversight on AI's technical sides, humans need to maintain a psychological sense of control within human-AI partnerships (Wen et al., 2022). This includes intervention mechanisms where humans require the ability to override AI decisions and actions when necessary. Well-defined authority boundaries can reduce decision conflicts in collaborative settings (Kumar & Thompson, 2024), so that clear delineation of decision and action rights between humans and AI systems is essential.

*Shared Responsibilities: Collaborative Decision-Making*

While establishing vertical leadership, shared responsibilities distribute decision-making responsibilities according to comparative advantages and different context. This approach recognizes that AI systems may possess superior capabilities in certain domains while humans excel in others. The concept of "human-in-the-loop" evolves to position humans as orchestrators who coordinate multiple AI systems and determine when and how to leverage AI capabilities (Dellermann et al., 2021). Recent research by Akata et al. (2024) also suggests that orchestrator frameworks, where humans strategically deploy AI systems based on their strengths, outperform both fully automated and traditionally supervised approaches.

(2) Principle 2: AI Empowering Humans

*Transformative Leadership: Augmentation Over Replacement*

The second principle focuses on how AI systems should enhance human capabilities rather than diminish or replace human contributions (Shneiderman, 2020). This transformative approach positions AI as an enabler of expanded human potential. Chiriatti et al. (2024) proposed that outsourcing certain cognitive tasks to AI constitutes a distinct psychological system, defined as 'System 0', forming an artificial, non-biological underlying layer of distributed intelligence that interacts with and augments both intuitive and analytical thinking processes, in parallel with Kahneman's System 1 (fast, intuitive thinking) and System 2 (slow, analytical thinking). Unlike these two systems, System 0, as an extension of human cognition, enhances and expands our thinking capabilities through dynamic interaction with AI. The empowerment forms human-AI collective intelligence (Willson & Daugherty, 2018), which can also transfer to humans' own knowledge development (Te'eni et al., 2023). However, AI's empowerment qualities can trigger identity loss, feelings of replaceability, and even existential fears about human obsolescence. Concerns about autonomous systems or AI losing control are also intensifying (Docherty, 2018).

Maintaining human irreplaceability is currently achieved through active alignment of AI with human values, as previously discussed. Effective human-AI collaboration requires systems that understand and align with human values and mental models (Russell, 2019). On one hand, instilling human values in AI facilitates shared mental models, with regulatory frameworks such as the European Union's AI Act and organizational ethical AI principles increasingly emphasizing value alignment (Jobin et al., 2019; Ghosh et al., 2025). On the other hand, developing AI systems with mental structures that align with human cognition forms team situational awareness more effectively. For example, AI systems trained to mirror human mental models (Murphy, 2024) and theory of mind capabilities (Rabinowitz et al., 2018) demonstrate superior efficiency in collaborative settings.

## 6. Application Analysis: Taking Autonomous Driving as an Example

Building upon the human-centered HAC framework and the four-level research agenda of HAC—encompassing team cognition, control, transaction, and relationship—autonomous driving emerges as a quintessential domain to examine these concepts in practice. In this context, the interaction between human drivers and autonomous systems necessitates a dynamic and context-sensitive leadership model, moving beyond static human control to a more fluid distribution of responsibility and a collaborative relationship.

6.1 Team Cognition Level

In autonomous driving, effective human-AI collaboration requires cognitive integration between drivers and intelligent systems. SMM allow both parties to develop unified representations of driving tasks and environments (Andrews et al., 2022). This cognitive alignment enables transformative leadership where AI enhances human learning while preserving human agency. Autonomous vehicles establish bidirectional recognition systems that monitor driver states while communicating system capabilities, creating robust TSA.

Social cognition manifests through interfaces that recognize driver emotional states and cognitive readiness (de Melo et al., 2021). These processes support shared leadership where decision authority shifts based on contextual advantages. Vehicles with theory of mind capabilities demonstrate superior collaboration in complex scenarios (Rabinowitz et al., 2018). Team decision-making represents a process where AI augments human judgment by providing analyzed information without diminishing driver authority (O'Neill et al., 2016).

6.2 Team Control Level

Function allocation exemplifies the vertical leadership structure necessary for human-centered collaboration. Autonomous systems implement dynamic allocation where control shifts based on situational demands and safety parameters. Contemporary systems adapt driving modes to environmental conditions while preserving driver override capabilities, maintaining human oversight and accountability. Adaptable automation—where drivers retain authority to modify automation levels—shows superior performance and acceptance compared to fully adaptive systems.

Action coordination involves synchronized behaviors between drivers and autonomous systems through both shared control and cooperative control. For example, drivers may allow the control on steering wheel from the autonomous systems for keeping the horizontal control of the vehicle steady, while longitudinal control may completely be controlled by the autonomous system. Various interaction modes emerge in driving scenarios: AI-guided interactions when vehicles request driver input; human-guided interactions when drivers correct system behaviors; and request-driven modes allowing selective engagement of autonomous feature. These different interaction modes reflect the fluid shared responsibility change.

6.3 Team Transaction Level

Communication pathways serve as critical conduits for implementing leadership dimensions. Efforts should be made to promote multi-channel natural interaction between the driver and the intelligent agent, actively exploring and researching effective human-machine interface design metaphors, paradigms, and cognitive architectures, such as cooperative cognitive interfaces to support human-machine cooperation in human-machine co-driving. Vehicles implementing the SA-based agent transparency model show improved performance, awareness, and trust without increasing workload (Chen et al., 2018; Mercado et al., 2016). Communication design must balance information completeness with cognitive accessibility, employing appropriate frequency, content relevance, and timing (Lyons et al., 2021). Interpretable interfaces that visualize vehicle trajectories demonstrate superior performance compared to text-based explanations, particularly in time-critical scenarios (Bhaskara et al., 2020).

6.4 Team Relationship Level

Trust dynamics represent fundamental relational processes determining collaboration quality. Following Gao et al.'s (2021) framework, driver trust evolves through dispositional, initial, ongoing, and post-task phases influenced by driver characteristics, system attributes, and situational variables. During the interaction, the driver obtains the AI system's intentions, reasoning logic, and system capability in a timely manner through the explainable and transparent cognitive interaction interface; The AI system can form trust to human driver by recognizing and understanding the user's intentions and emotional states by monitoring the driver's multimodal information. Based on this, the driver and the autonomous system form a human-machine relationship of mutual trust, which will also be crucial for the different levels of leadership to take effect.

6.5 Human-Centeredness

For human-centeredness, the involvement of human in the loop has already been manifested through the previous four level, where human and the autonomous system collaborate on achieving team understanding and team control. The ultimate control of human manifests through design choices prioritizing driver agency and control. Effective autonomous systems enhance capabilities rather than replace drivers, supporting through cognitive augmentation while preserving human decision authority. This approach recognizes complementary capabilities: AI excels at continuous monitoring and precise control, while humans retain superior contextual understanding and ethical judgment (Willson & Daugherty, 2018). The resulting relationship achieves mutual trust, balanced authority, and collaborative harmony essential for successful human-AI collaboration in safety-critical domains.

## 7. Conclusion

This chapter systematically depicts the evolving human-machine relationship in the context of the intelligent era, identifying HAC as a novel form of interaction introduced by autonomous intelligent machines. With the advent of intelligent systems, the challenge is no longer whether humans and AI can collaborate—but how this collaboration can be structured to ensure human leadership, agency, and ethical integrity. This chapter has argued for a foundational reorientation of HAC: away from a technology-centric view and toward a human-centered, human-led paradigm. In such a framework, intelligent systems are designed not only for task efficiency, but for enhancing human experience, respecting autonomy, and fostering meaningful partnerships. This transformation necessitates the adoption of new research paradigms and orientations to effectively address the challenges posed by human-centered HAC. Building on this foundation, the chapter provides a comprehensive summary of existing research methodologies, encompassing research paradigms, platforms, and variables, as well as the key areas of study within the HAC framework, including the team cognition, control, transaction and relationship. Additionally, the chapter proposes the conceptual framework of HCHAC for human-centered HAC, and presents autonomous driving as a case study to exemplify the practical applications of HAC and the human-centered framework.

Although numerous studies on HAC have been conducted over the past ten years, laying a foundational basis for this chapter, this domain remains in its early stages. To achieve efficient, safe, and HCHAC, several critical issues must be addressed in future research:

(1) Current HAC research often borrows concepts from human-human teaming and traditional human-machine interaction studies. However, HAC is distinct from these domains, necessitating the exploration of unique theories and the establishment of performance evaluation systems tailored specifically for HAC.

(2) Future research HAC should prioritize the currently underexplored area of human leadership within these teams. For example, future research needs to define clear leadership roles and styles appropriate for AI-integrated environments, design AI systems that can recognize, support, and adapt to human leadership, and develop robust metrics to evaluate leadership effectiveness. Training programs and simulations should be expanded to include human leadership scenarios, preparing users to effectively guide AI systems in dynamic and high-stakes settings.

(3) HAC requires mutual trust between humans and machines, shared SA and decision-making authority, but there is currently a lack of empirically testable specific HAC theoretical models. Future research should focus on developing bidirectional SA, trust, and decision-making conceptual and computational models that encapsulate HAC characteristics.

(4) As AI technology develops, HAC not only happens in mechanical tasks but also occurs in complex social contexts. This poses new requirements for HAC, particularly regarding situational awareness and mental models, which must consider more complex social information and macro-social factors. Currently, HAC in complex social contexts has not received sufficient attention and urgently needs in-depth exploration.

(5) Function allocation based on HAC requires adaptability and adaptability, but the rules for using adaptability and adaptability are not clear, and it is necessary to systematically discuss this issue in the future.

(6) The interaction interface between humans and intelligent entities is extremely important for humans to understand and trust intelligent entities. Some researchers currently position this interface as a cooperative cognitive interface (You et al., 2022), but the cooperative cognitive interface is more conceptual, and it is necessary to further explore the interface properties truly applicable to human-intelligence interaction in the future, and develop new paradigms and models based on this new interface for testing.

(7) With the rapid advancement of AI technology, intelligent entities have surpassed humans in some aspects, and human abilities are limited by evolution and will remain relatively stable in the short term. Therefore, future research needs to start from the perspective of human factors science, considering the nature of tasks and human characteristics, actively participating in the research and design of intelligent entity behavior, cognition, and social cognitive abilities, in order to enhance the level of collaborative cooperation in HAC and achieve natural interaction.